\documentclass[twocolumn,prb,superscriptaddress,preprintnumbers,amsmath,amssymb,showpacs,floatfix]{revtex4}

\usepackage{graphicx}
\usepackage{dcolumn}
\usepackage{bm}
\usepackage{color}

\begin{document}

\title{Epitaxial and layer--by--layer growth of EuO thin films on yttria--stabilized cubic zirconia (001) using MBE distillation}

\author{R. Sutarto}
 \affiliation{ II. Physikalisches Institut, Universit\"{a}t zu K\"{o}ln,
  Z\"{u}lpicher Str. 77, 50937 K\"{o}ln, Germany}
\author{S. G. Altendorf}
 \affiliation{ II. Physikalisches Institut, Universit\"{a}t zu K\"{o}ln,
 Z\"{u}lpicher Str. 77, 50937 K\"{o}ln, Germany}
\author{B. Coloru}
 \affiliation{ II. Physikalisches Institut, Universit\"{a}t zu K\"{o}ln,
 Z\"{u}lpicher Str. 77, 50937 K\"{o}ln, Germany}
\author{M. Moretti Sala}
 \affiliation{ II. Physikalisches Institut, Universit\"{a}t zu K\"{o}ln,
 Z\"{u}lpicher Str. 77, 50937 K\"{o}ln, Germany}
\author{T. Haupricht}
 \affiliation{ II. Physikalisches Institut, Universit\"{a}t zu K\"{o}ln,
 Z\"{u}lpicher Str. 77, 50937 K\"{o}ln, Germany}
\author{C.~F.~Chang}
 \affiliation{ II. Physikalisches Institut, Universit\"{a}t zu K\"{o}ln,
 Z\"{u}lpicher Str. 77, 50937 K\"{o}ln, Germany}
\author{Z.~Hu}
 \affiliation{ II. Physikalisches Institut, Universit\"{a}t zu K\"{o}ln,
 Z\"{u}lpicher Str. 77, 50937 K\"{o}ln, Germany}
\author{C. Sch{\"u}{\ss}ler-Langeheine}
 \affiliation{ II. Physikalisches Institut, Universit\"{a}t zu K\"{o}ln,
 Z\"{u}lpicher Str. 77, 50937 K\"{o}ln, Germany}
\author{N. Hollmann}
 \affiliation{ II. Physikalisches Institut, Universit\"{a}t zu K\"{o}ln,
 Z\"{u}lpicher Str. 77, 50937 K\"{o}ln, Germany}
\author{H. Kierspel}
 \affiliation{ II. Physikalisches Institut, Universit\"{a}t zu K\"{o}ln,
 Z\"{u}lpicher Str. 77, 50937 K\"{o}ln, Germany}
\author{H. H. Hsieh}
 \affiliation{Chung Cheng Institute of Technology, National Defense University, Taoyuan 335, Taiwan}
\author{H.-J. Lin}
 \affiliation{National Synchrotron Radiation Research Center, 101 Hsin-Ann Road, Hsinchu 30077, Taiwan}
\author{C. T. Chen}
 \affiliation{National Synchrotron Radiation Research Center, 101 Hsin-Ann Road, Hsinchu 30077, Taiwan}
\author{L. H. Tjeng}
 \affiliation{ II. Physikalisches Institut, Universit\"{a}t zu K\"{o}ln,
 Z\"{u}lpicher Str. 77, 50937 K\"{o}ln, Germany}

\date{\today}

\begin{abstract}
We have succeeded in growing epitaxial and highly stoichiometric
films of EuO on yttria--stabilized cubic zirconia (YSZ) (001). The
use of the Eu--distillation process during the molecular beam
epitaxy assisted growth enables the consistent achievement of
stoichiometry. We have also succeeded in growing the films in a
layer--by--layer fashion by fine tuning the Eu vs. oxygen
deposition rates. The initial stages of growth involve the limited
supply of oxygen from the YSZ substrate, but the EuO stoichiometry
can still be well maintained. The films grown were sufficiently
smooth so that the capping with a thin layer of aluminum was leak
tight and enabled \textit{ex situ} experiments free from trivalent
Eu species. The findings were used to obtain recipes for better
epitaxial growth of EuO on MgO (001).
\end{abstract}

\pacs{68.55.-a, 75.70.Ak, 78.70.Dm, 79.60.Dp, 81.15.Hi}

\maketitle

\section{Introduction}
Stoichiometric EuO is a ferromagnetic semiconductor with a Curie
temperature ($T_C$) of 69~K and a band gap of about 1.2 eV at room
temperature.\cite{mauger86a} Upon electron doping, the material
shows a wealth of spectacular phenomena, including a
metal--to--insulator transition and colossal magnetoresistance,
where the change in resistivity can exceed 8--10 orders of
magnitude.\cite{oliver72a,shapira73a,shapira73b} The Curie
temperature can also be doubled by electron doping\cite{mauger78}
and even almost tripled by pressure.\cite{dimarzio87a,elmeguid90a}
In the ferromagnetic state the conduction band shows a splitting
of about 0.6~eV between the spin--up and spin--down states leading
to an almost 100\% spin polarization of the charge carriers in
electron doped EuO.\cite{steeneken02a} These properties make EuO a
very attractive candidate for fundamental research in the field of
spintronics.

EuO in thin film form has been studied already in the late 1960s
and early 1970s. Use was made of coevaporation of Eu and
Eu$_2$O$_3$,\cite{ahn67a,ahn68a,ahn70a,lee70a,lee71a,ahn71a,mcguire71a,suits71a,suits71b}
and later, of evaporation of Eu in an oxygen
atmosphere,\cite{paparoditis71a,llinares73a,llinares73b,massenet74a,llinares75a}
all under \textit{technical} vacuum conditions, i.e., pressures in
the range of 10$^{-6}$--10$^{-5}$ mbar. The preparation of thin
films was considered as a convenient alternative synthesis route
for EuO, alternative to that of the bulk synthesis which required
very high temperatures with a delicate phase
diagram.\cite{schafer72a} The use of the thin--film preparation
route specifically facilitated doping dependence studies using
rare--earth and transition--metal
impurities.\cite{ahn68a,lee71a,ahn71a,mcguire71a,suits71a}

After a pause of two decades, a strong renewed interest in EuO
thin films emerged in recent years. This time it is the thin--film
community which is making the attempt to utilize the extraordinary
properties of EuO for device
applications.\cite{steeneken02a,roesler94a,sohma97a,iwata00a,iwata00b,
steeneken02b,lettieri03a,matsumoto04a,holroyd04a,santos04a,ott06a,
negusse06a,lee07a,schmehl07a,muhlbauer08a,laan08a,ingle08a,ulbricht08a,panguluri08a}
Part of the motivation also originates from the fact that
tremendous progress has been made in preparation technologies,
e.g., molecular beam epitaxy (MBE) under ultra--high--vacuum
conditions, and that new analysis methods have become available,
e.g., synchrotron--based spectroscopies. All in all, these new
efforts have culminated in a high point such that EuO thin films
can be grown epitaxially on Si, demonstrating its potential for
spintronics applications.\cite{schmehl07a}

It is remarkable in the recent EuO research
\cite{roesler94a,sohma97a,iwata00a,iwata00b,
lettieri03a,matsumoto04a,holroyd04a,santos04a,
negusse06a,lee07a,schmehl07a,muhlbauer08a,laan08a,ulbricht08a,panguluri08a}
that control of the stoichiometry is nevertheless still a serious
issue. Many studies reported that Eu$^{3+}$ ions were present in
their films and/or that the magnetic moment per f.u. was not close
to the expected 7$\mu_B$ for a $4f^7$ system. It is not clear in
what precision the relative supply rates of oxygen and europium
were controlled in these works. We will show below that this
control need not be precise as long as one is in the so--called
Eu--distillation condition during growth.

Also layer--by--layer growth has -- to our knowledge -- never been
mentioned, although epitaxy has been often reported. We therefore
set out to do a renewed growth study. We have chosen for
yttria--stabilized cubic zirconia (YSZ) as
substrate\cite{ingel86a,yashima94a}: the lattice constant of YSZ
is 5.142~\AA, practically identical to the 5.144~\AA\ value for
EuO at room temperature,\cite{henrich94a} and epitaxy has been
reported already.\cite{roesler94a,steeneken02b,schmehl07a} Yet, it
was also claimed that control stoichiometry is extremely
difficult,\cite{steeneken02b,schmehl07a} related to the fact that
YSZ acts as a source of oxygen\cite{minh93a} during the MBE growth
process. We will show below that we do have achieved full control
of the growth process, i.e., with stoichiometry control and
layer--by--layer epitaxial growth, and that we can use these
results as a firm basis for further studies including the doping
dependence and the specific use of well--defined interfaces.

\begin{figure*}[t]
\includegraphics*[width = 17 cm] {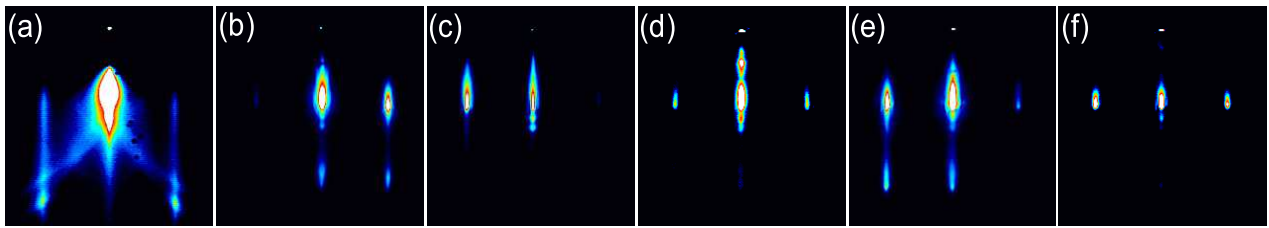}
\caption{\label{Fig_1} (Color online) RHEED photographs of (a)
clean and annealed YSZ (001), [(b)--(d)] EuO films on YSZ (001)
after 10 min of growth at $400^{\circ}$C with 8.1--8.2~\AA/min Eu
flux rates, and [(e)--(f)] with 4.2--4.3~\AA/min Eu flux rates.
The oxygen pressure in the chamber was $4\times10^{-8}$ mbar for
(b), and $2\times10^{-8}$ mbar for (c) and (e). No oxygen was
supplied into the chamber for (d) and (f). The RHEED electron
energy was 20 keV with the beam incident along the [100]
direction.}
\end{figure*}

\section{Experiment}
The EuO films were grown in an ultra--high--vacuum MBE facility
with a base pressure of 2$\times$10$^{-10}$ mbar, maintained by a
cryopump. High purity Eu metal from AMES Laboratory was evaporated
from an EPI effusion cell with a BN crucible at temperatures
between 460$^{\circ}$C and 525$^{\circ}$C. Proper degassing of the
Eu material (mostly hydrogen gas) ensured that during Eu
evaporation the pressure was kept below 3$\times$10$^{-9}$ mbar.
The Eu deposition rate (4--8 \AA/min) was calibrated using a
quartz--crystal monitor which was moved to the sample growth
position prior and after each growth. Molecular oxygen was
supplied through a leak valve, and its pressure
(4--16$\times$10$^{-8}$ mbar) was monitored using an ion--gauge
and a mass--spectrometer. The growth was terminated by closing
first the oxygen leak valve and then the Eu shutter after 30 s.

As substrates we used epipolished single crystals of YSZ from
SurfaceNet GmbH and cleaved single crystals of MgO from
TBL--Kelpin. The surface normal of the substrates are all the
(001). The lattice constant of YSZ is 5.142~\AA, very close to the
5.144~\AA\ value for EuO at room temperature. The lattice constant
of MgO is 4.21 \AA.\cite{henrich94a} Prior to growth the
substrates were annealed \textit{in situ} at $T=600^{\circ}$C in
an oxygen atmosphere of $5\times10^{-7}$ mbar for at least 120 min
in the case of YSZ, and in $1\times10^{-7}$ mbar for at least 60
min for MgO, in order to obtain clean and well--ordered substrate
surfaces. The substrates were kept at $T=400^{\circ}$C during
growth.

The MBE facility is supplied with the EK--35--R reflection
high--energy electron diffraction (RHEED) system from STAIB
Instruments for \textit{in situ} and \textit{online} monitoring of
the growth. The MBE facility is attached to an ultra--high--vacuum
$\mu$--metal photoemission chamber equipped with a Scienta
SES--100 electron energy analyzer and a Vacuum Generators twin
crystal monochromatized Al--$K_{\alpha}$ ($h\nu$ = 1486.6~eV)
source for \textit{in situ} x--ray photoelectron spectroscopic
(XPS) analysis. The overall energy resolution was set to 0.35 eV
and the Fermi level $E_F$ was calibrated using a polycrystalline
Au reference. The $\mu$--metal chamber is also equipped with a
Vacuum Generators Scientific T191 rear--view low--energy electron
diffraction (LEED) system for further \textit{in situ} structural
characterization. The base pressure of the $\mu$--metal chamber is
1$\times$10$^{-10}$ mbar, and all characterizations herein were
carried out at room temperature.

The MBE facility is also connected to a separate
ultra--high--vacuum chamber for the evaporation of aluminum as
protective capping layer of the air--sensitive EuO films. This
allows the \textit{ex situ} characterizations using x--ray
reflectivity (XRR), superconducting quantum interference device
(SQUID), and x--ray absorption spectroscopy (XAS). The thickness
of the aluminum capping is about 20--40 \AA. The XRR measurements
were carried out using a Siemens D5000 diffractometer. The
magnetic properties of the films were determined using a Quantum
Design MPMS--XL7 SQUID magnetometer. The XAS measurements were
performed at the Dragon beamline of the National Synchrotron
Radiation Research Center (NSRRC) in Taiwan. The spectra were
recorded using the total--electron--yield method and the
photon--energy resolution at the Eu $M_{4,5}$ edges ($h\nu \approx
1100$--1180~eV) was set at $\approx$ 0.6~eV.

\section{\label{three} Initial stages of growth on YSZ}
\begin{figure}[t]
\includegraphics*[scale = 0.3] {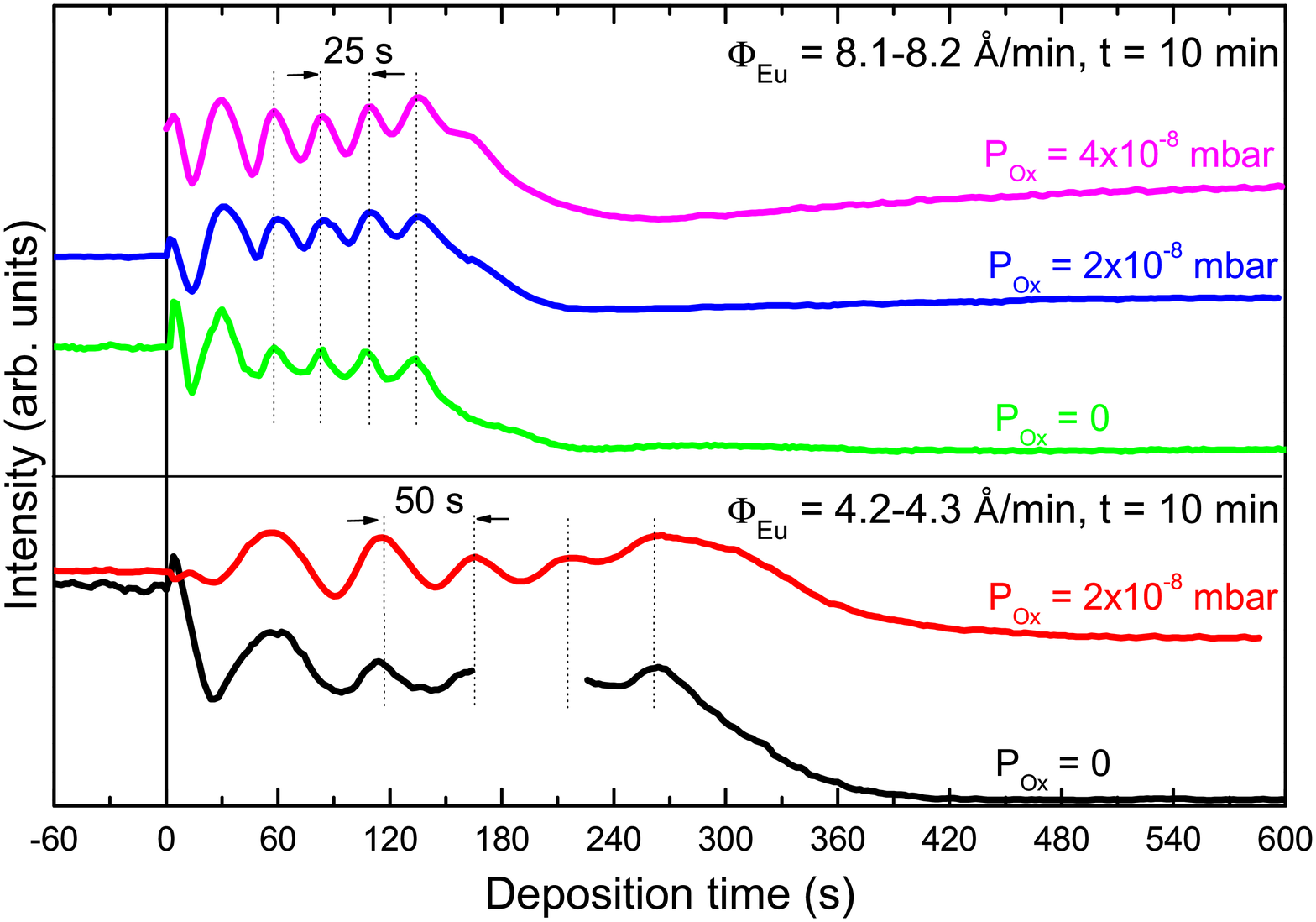}
\caption{\label{Fig_2} (Color online) RHEED intensity oscillations
of the specularly reflected electron beam, recorded during the
deposition of EuO films on YSZ (001) using oxygen pressures
($P_{\rm Ox}$) and Eu flux rates ($\Phi_{\rm Eu}$) as indicated.
The corresponding RHEED photographs after 10 min of growth are
displayed in Figs.~1(b)--1(f).}
\end{figure}

Figure~1(a) shows the RHEED photograph of the clean and annealed
YSZ (001) before growth, and Figs.~1(b)--1(e) show the photographs
after 10 min of EuO growth. The Eu flux rates were
8.1--8.2~\AA/min for (b)--(d) and 4.2--4.3~\AA/min for (e)--(f).
The oxygen pressure in the chamber was $4\times10^{-8}$ mbar for
(b) and $2\times10^{-8}$ mbar for (c) and (e). No oxygen was
supplied into the chamber for (d) and (f). The YSZ substrate
temperature was kept at $T=400^{\circ}$C during growth. The
important result is that the general features of the RHEED
patterns did not change during growth and that they are very
similar to those of the clean YSZ for all Eu and O growth
conditions. The distance between the streaks of the EuO films is
identical to that of the pure YSZ, confirming that the in--plane
lattice constants of EuO and YSZ are very closely matched.

Figure~2 shows the time dependence of the RHEED intensity of the
specularly reflected beam during the EuO growth. We can clearly
observe oscillations which are indicative for a two--dimensional
(2D) layer--by--layer or Frank--van der Merwe growth mode. It is
surprising that there are only five to six oscillations for all
deposition conditions as indicated in Fig.~2 and that these
oscillations even exist in the absence of oxygen in the MBE
chamber. It is important to note that the oscillation period does
\textit{not} depend on the oxygen pressure $P_{\rm Ox}$, thus also
in the case of no oxygen in the chamber. This indicates that the
oxygen needed for the formation of EuO must also come from the YSZ
substrate. The $T=400^{\circ}$C substrate temperature apparently
provides sufficient mobility for the oxygen ions to migrate to
form at least five or six EuO layers. The oscillation period,
which represents a formation of a new atomic single layer, is
determined only by the Eu flux rate $\Phi_{\rm Eu}$: reducing it
by a factor of 2, from 8.1--8.2 to 4.2--4.3~\AA/min, doubles the
period, from 25 to 50 s.

\begin{figure}[t]
\includegraphics*[scale = 0.3] {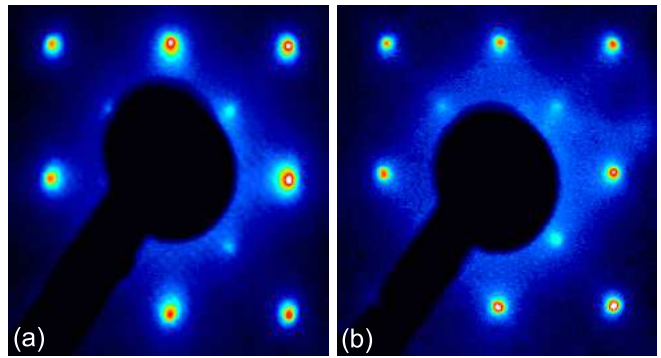}
\caption{\label{Fig_3} (Color online) LEED photographs of
epitaxial EuO films on YSZ substrate, grown for 10 min at
$400^{\circ}$C in the absence of oxygen in the MBE chamber using
(a) a 8.2~\AA/min Eu flux rate and recorded at electron beam
energy of 215 eV, and (b) a 4.3~\AA/min Eu flux rate and recorded
at electron beam energy of 213 eV.}
\end{figure}

\begin{figure*} [t]
\includegraphics*[width = 17 cm] {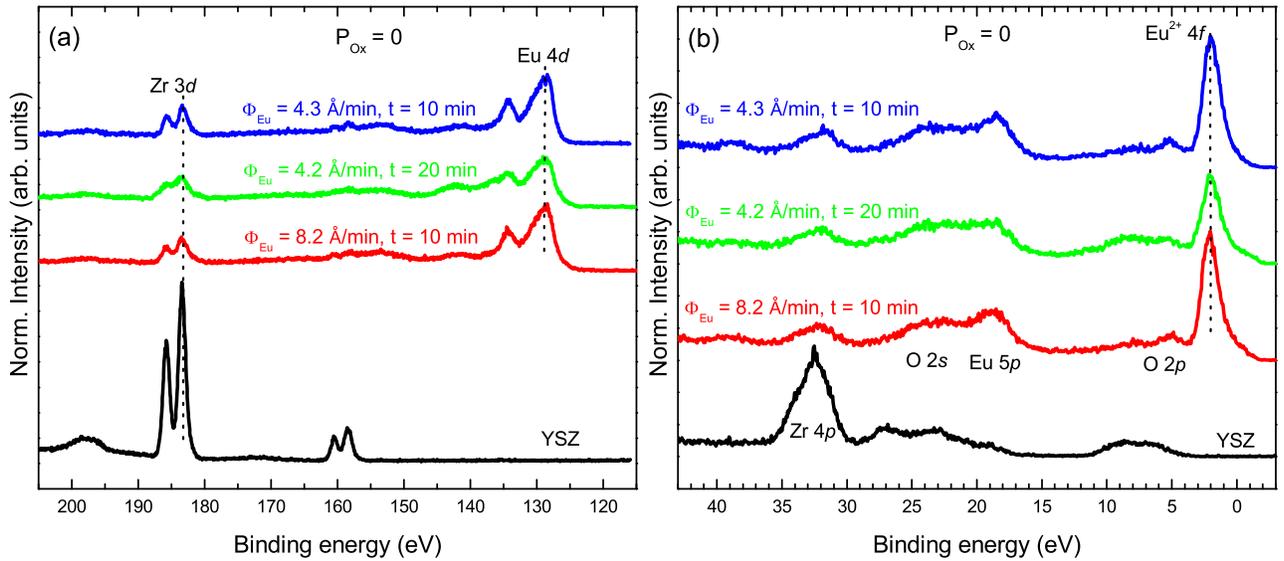}
\caption{\label{Fig_4} (Color online) (a) Zr~$3d$~--~Eu~$4d$
core~level XPS spectra and (b) Zr~4$p$~--~O~2$s$~--~Eu~5$p$
core~level and O~2$p$~--~Eu~4$f$ valence~band XPS spectra of EuO
films on YSZ (001), grown at $400^{\circ}$C in the absence of
oxygen in the MBE chamber. The spectra were collected at normal
emission. From top to bottom: EuO film after 10 min of growth
using a 4.3~\AA/min Eu flux rate, after 20 min using 4.2~\AA/min,
after 10 min using 8.2~\AA/min, and clean YSZ substrate.}
\end{figure*}

LEED photographs for all these films displayed a good single
crystallinity. Figure~3 depicts examples for the case of no oxygen
in the MBE chamber during growth. Also here we can observe a
perfect (001) surface of the EuO rock--salt structure, consistent
with the RHEED results. The LEED photographs were taken at
electron beam energies of 213--215~eV since lower energies did not
provide stable patterns due to charging.

To investigate the implications of observing only five to six
oscillations, we also carried out photoemission experiments on
those films. Figure~4, left panel (a), shows the Zr~$3d$ and
Eu~$4d$ core level XPS spectra which were collected at normal
emission. It can be clearly seen that the Zr signal is reduced
when comparing the clean YSZ (bottom curve -- black) with the
EuO--covered YSZ (top three curves -- blue, green, and red). The
EuO films here were grown without oxygen in the MBE chamber.
Different Eu flux rates and total time of growth are indicated in
the figure. It is remarkable that the EuO--covered YSZ spectra
have very similar Zr signals, and also equal Eu intensity, despite
the fact that the total amount of Eu exposure is twice as large in
the two middle curves (green and red) than in the top one (blue).
This indicates that in the absence of oxygen in the MBE chamber,
the growth of EuO is limited to five to six monolayers only and
that the rest of the deposited Eu metal is re--evaporated back
into the vacuum (the substrate temperature is $400^{\circ}$C). In
other words, the sticking coefficient for Eu after the completion
of five to six monolayers is reduced to zero, suggesting that
oxygen transport through EuO is much more difficult than in YSZ.

\begin{figure}[t]
\includegraphics*[scale = 0.3] {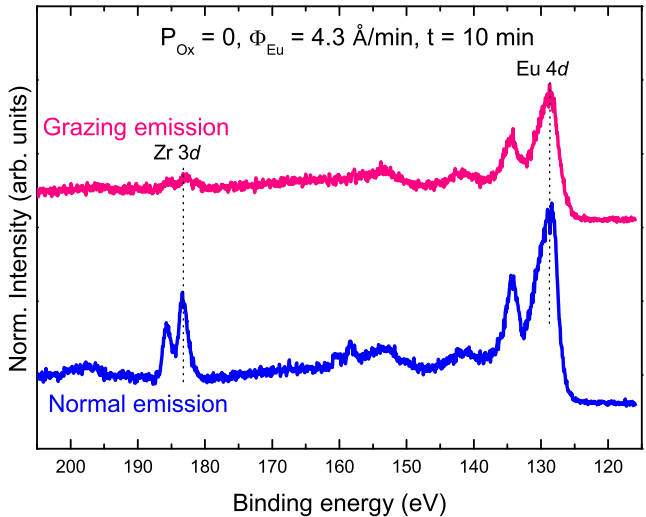}
\caption{\label{Fig_5} (Color online) Take--off angle dependence
of the Zr~$3d$~--~Eu~$4d$ core~level XPS spectra of a EuO film on
YSZ (001). Top: grazing emission, i.e., $\Theta=70^{\circ}$ with
respect to the surface normal. Bottom: normal emission. The film
was grown at $400^{\circ}$C for 10 min with a 4.3~\AA/min Eu flux
rate in the absence of oxygen in the MBE chamber.}
\end{figure}

\begin{figure*}[t]
\includegraphics*[width = 17 cm] {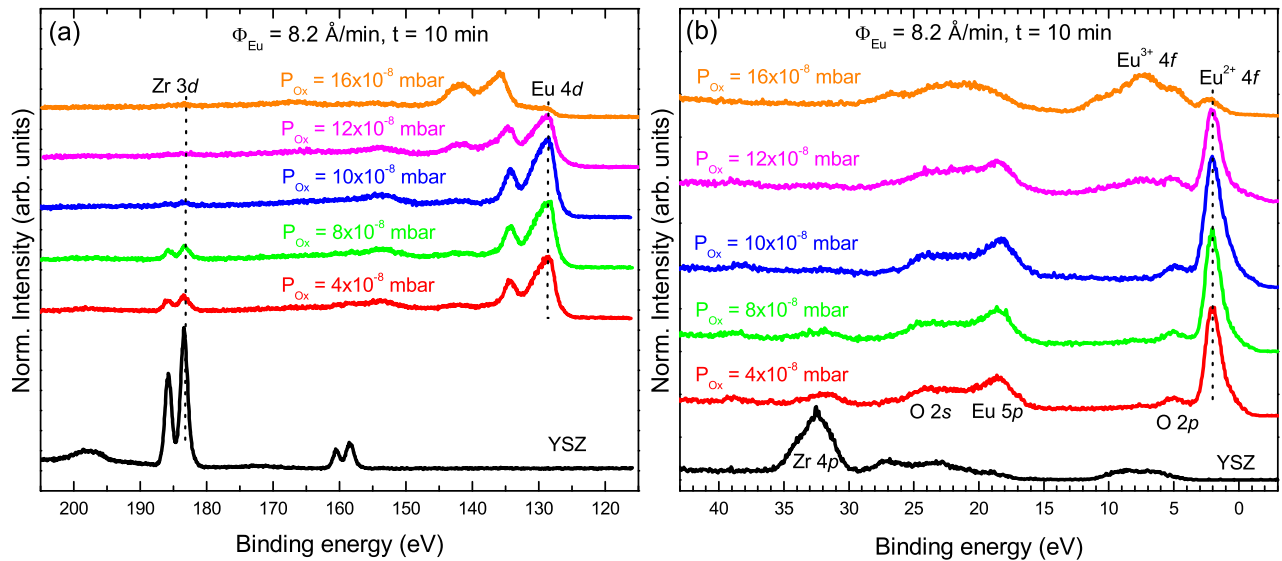}
\caption{\label{Fig_6} (Color online) (a) Zr~$3d$~--~Eu~$4d$
core~level XPS spectra and (b) Zr~4$p$~--~O~2$s$~-~Eu~5$p$
core~level and O~2$p$~--~Eu~4$f$ valence~band XPS spectra of EuO
films on YSZ (001), grown at $400^{\circ}$C with a 8.2~\AA/min Eu
flux rate for 10 min. The spectra were collected at normal
emission. From top to bottom: EuO films grown under oxygen
pressures of 16, 12, 10, 8, and $4\times10^{-8}$ mbar, and clean
YSZ substrate.}
\end{figure*}

\begin{figure}[t]
\includegraphics*[scale = 0.3] {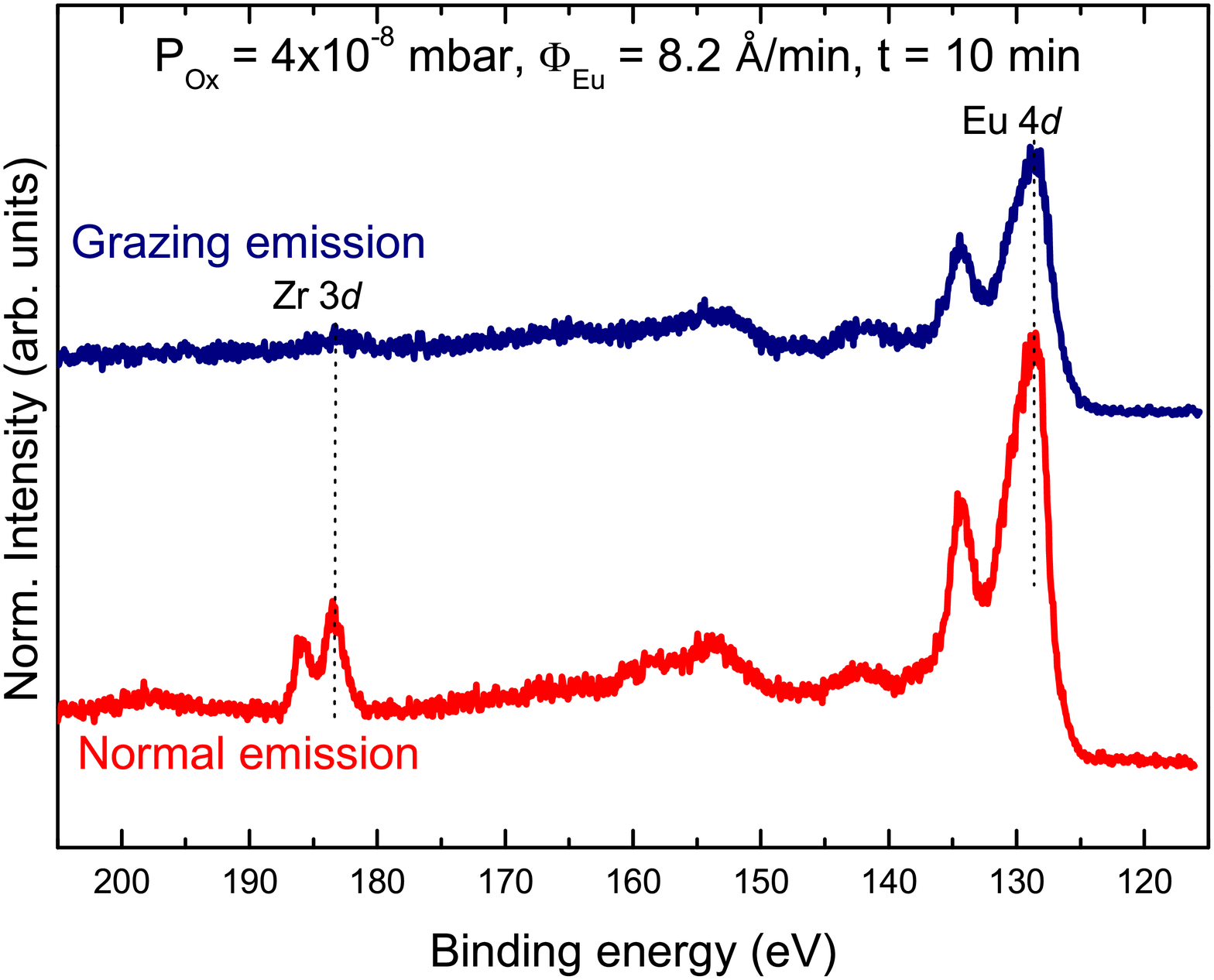}
\caption{\label{Fig_7} (Color online) Take--off angle dependence
of the Zr~$3d$~--~Eu~$4d$ core~level XPS spectra of a EuO film on
YSZ (001). Top: grazing emission, i.e., $\Theta=70^{\circ}$ with
respect to the surface normal. Bottom: normal emission. The film
was grown at $400^{\circ}$C under a $4\times10^{-8}$ mbar oxygen
pressure and a 8.2~\AA/min Eu flux rate for 10 min.}
\end{figure}

It is also important to investigate the chemical state of the Eu.
Figure~4, right panel (b), depicts the O~$2p$ and Eu~$4f$ valence
band spectra together with the Zr~$4p$, O~$2s$, and Eu~$5p$ core
levels. The Eu $4f$ lineshape in all the films is very
characteristic for a Eu$^{2+}$ system. The multiplet structure
typical for Eu$^{3+}$ is not visible. One can also observe that
the O $2p$ spectrum at 6--10 eV binding energy for YSZ is
converted into the O $2p$ valence band at 4--7 eV typical for EuO.
\cite{steeneken02a} All these demonstrate that only EuO has been
formed, free from Eu$_2$O$_3$ or Eu$_3$O$_4$ contaminants. This
also means that YSZ can only oxidize Eu into the 2+ state, and
definitely not into the 3+.

We have also carried out take--off angle--dependent XPS
experiments on the films. Figure~5 shows the Zr~$3d$ and Eu~$4d$
core level XPS spectra of one of the EuO films of Fig.~4~(a)
collected at grazing emission, i.e., $\Theta=70^{\circ}$ with
respect to the surface normal, and at normal emission. One can
clearly see that the Eu signal is not significantly reduced but
the Zr signal has almost disappeared in the grazing--emission
geometry. Since grazing emission means more surface sensitivity,
this result confirms not only that the EuO film is on top of the
YSZ substrate with negligible intermixing of the cations but also
that the film is closed and flat.

Figure~6 depicts the Zr~$3d$ and Eu~$4d$ core level spectra (left
panel), and the Zr~$4p$, O~$2s$, and Eu~$5p$ core level together
with O~$2p$ and Eu~$4f$ valence band spectra (right panel) of EuO
films grown with the supply of oxygen in the MBE chamber. Various
oxygen pressures have been used as indicated in the figure. The Eu
flux rate and the deposition time are identical for these films.
One can clearly observe that the Zr~$3d$ signal is getting smaller
when the oxygen pressure is increased, indicating that the
thickness of the EuO film becomes larger. The lineshapes of the
Eu~$4d$ and Eu~$4f$ levels are those of divalent Eu for pressures
up to 12$\times$10$^{-8}$ mbar. For a pressure of
12$\times$10$^{-8}$ mbar or higher, however, the Eu~$4d$ and
Eu~$4f$ spectral shapes start to change and show characteristics
which indicate the presence of trivalent Eu. Apparently, for a Eu
flux rate of 8.2~\AA/min and substrate temperature of
$400^{\circ}$C, 10--12$\times$10$^{-8}$ mbar is the critical
oxygen pressure below which EuO films can be made on YSZ (001)
free from any Eu$_2$O$_3$ or Eu$_3$O$_4$ type of impurity phases.

We have also measured the photoemission spectra of these EuO films
under grazing take--off angle conditions. Again, the Eu $4d$ core
level and Eu $4f$ valence band spectra are all 2+ as long as the
oxygen pressures in the MBE chamber are below the critical
10--12$\times$10$^{-8}$ mbar value. One example is shown in
Fig.~7, where an oxygen pressure of $4\times10^{-8}$ mbar was
used. This figure demonstrates that also the surface region of the
films is free from Eu$^{3+}$ species. It is interesting to compare
the grazing with the normal emission spectra and also with the
spectra displayed in Fig.~5. One can clearly see in Fig.~7 that
the Zr~$3d$ signal is very much suppressed in grazing emission,
even more suppressed than in the grazing emission spectrum of
Fig.~5. This shows that the EuO films grown with the supply of
oxygen in the MBE chamber are thicker, a not so surprising and yet
very consistent observation since without oxygen we have found
that the EuO film growth is limited to five to six monolayers.

The following picture can now be drawn about the initial stages of
growth of EuO film on YSZ (001). In case there is no oxygen in the
MBE chamber, YSZ is supplying all the oxygen required to form EuO.
The film is perfectly free from Eu$^{3+}$ species. In case oxygen
is present in the MBE chamber, YSZ is supplying only the amount of
oxygen that is needed to complete the formation of EuO. This also
explains why the growth rate of the first five to six monolayers
is determined only by the Eu flux rate and is totally independent
of the supply of oxygen pressure in the MBE chamber, see Fig.~2.
It is important that the pressure is kept below the critical value
of 10--12$\times$10$^{-8}$ mbar as we will discuss in more detail
in Sec.~\ref{four}.

Based on the comprehensive set of RHEED, LEED, and XPS data,
including the RHEED intensity oscillations, we have now
demonstrated that EuO thin films can be grown epitaxially in a
layer--by--layer fashion with good control of its chemical state.
The supply of oxygen from the YSZ does not do any harm, and in
fact, it can be utilized as a welcoming method to calibrate the Eu
flux rate accurately, e.g., the 8.1--8.2 \AA/min from the quartz
crystal monitor corresponds to the growth of one monolayer EuO per
25 s.

\section{\label{four} Sustained growth of E\lowercase{u}O on YSZ}
\begin{figure}[t]
\includegraphics*[scale = 0.3] {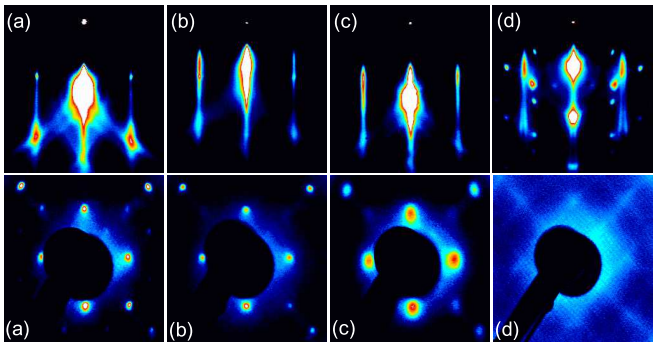}
\caption{\label{Fig_8} (Color online) Top panels: RHEED
photographs of EuO films on YSZ (001) grown at $400^{\circ}$C with
8.0--8.3~\AA/min Eu flux rates under oxygen pressures of -- from
left to right -- (a) 4, (b) 8, (c) 10, and (d) $12\times10^{-8}$
mbar. The deposition times were 200, 200, 100, and 100 min,
respectively. The RHEED electron energy was 20 keV with the beam
incident along the [100] direction. Bottom panels: corresponding
LEED photographs. The LEED electron energies were 369, 368, 370,
and 266 eV, respectively.}
\end{figure}

\begin{figure}[t]
\includegraphics*[scale = 0.3] {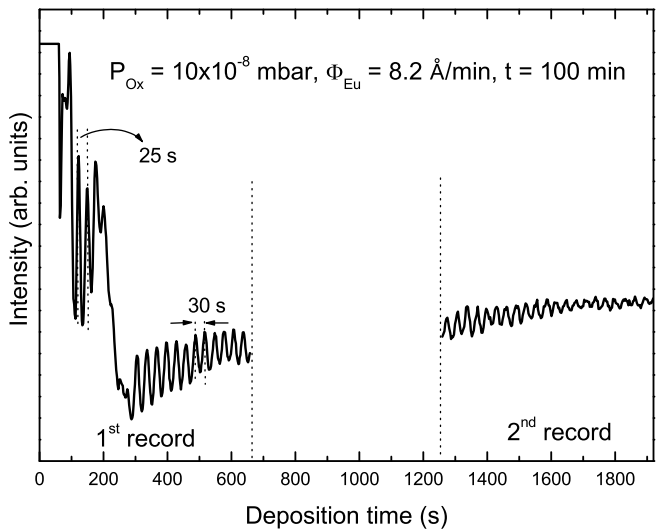}
\caption{\label{Fig_9} RHEED intensity oscillations of the
specularly reflected electron beam, recorded during deposition of
a EuO film on YSZ (001) grown at $400^{\circ}$C using a
$10\times10^{-8}$ mbar oxygen pressure and a 8.2~\AA/min Eu flux
rate.}
\end{figure}

\begin{figure}[t]
\includegraphics*[scale = 0.3] {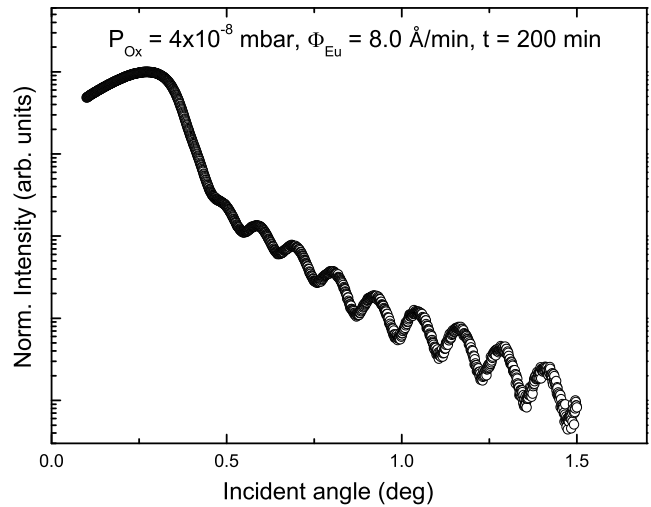}
\caption{\label{Fig_10} XRR curve of epitaxial EuO film on YSZ
(001) grown for 200 min at $400^{\circ}$C using a $4\times10^{-8}$
mbar oxygen pressure and a 8.0~\AA/min Eu flux rate.}
\end{figure}

Having shown that the initial stages of growth of EuO on YSZ (001)
can be made quite perfect, we now investigate whether thicker EuO
films can be prepared while keeping the epitaxy and, especially,
the layer--by--layer growth mode. We therefore have grown films
for longer deposition times, e.g., between 100 and 200 min, using
a series of finely intervalled pressures for the oxygen, e.g., 4,
8, 10, 12, and 16 $\times10^{-8}$ mbar. The Eu flux rates were
kept at 8.0--8.3~\AA/min. The resulting thickness of the films
varies between roughly 300 and 800~\AA\ as will be discussed
later. The RHEED and LEED results are plotted in Fig.~8. One can
clearly see that excellent epitaxial growth has been achieved for
(a) 4, (b) 8, and (c) 10 $\times10^{-8}$ mbar oxygen pressures.
For 12 $\times10^{-8}$ mbar (d) or higher pressures, however, the
appearance of additional spots in the RHEED indicates that the
surface structure starts to change, and the absence of a pattern
in the LEED even suggests appreciable surface roughness. The LEED
photographs were taken at electron beam energies of 368--370~eV
since lower energies did not provide stable patterns due to
charging.

As discussed in Sec~\ref{three}, the initial five to six
oscillations of the specular reflected RHEED beam intensity are a
unique feature for the initial stages of EuO growth on the YSZ
(001). These initial oscillations do always occur, i.e.,
independent of the oxygen pressure in the MBE chamber, unless the
pressure exceeds a critical value above which trivalent Eu species
are formed. Remarkable is that no more oscillations can be
observed beyond these five to six when growing thicker films. This
is the case for a wide range of oxygen pressures. There is one
exception: for a pressure of $10\times10^{-8}$ mbar, we were able
to see further RHEED intensity oscillations. Figure~9 shows that
the initial five to six oscillations are then followed by at least
50 more oscillations. It demonstrates that a layer--by--layer
growth mode for EuO is possible during the sustained growth.
Interestingly, the oscillation period during the sustained growth
is similar and yet a little bit larger than during the initial
stages of growth: 30 s against 25 s. Apparently, the oxygen
pressure must be close to and yet a little less than the critical
value in order to maintain the layer--by--layer growth mode, while
the Eu flux determines the growth rate in the initial stages, i.e.
8.0--8.2~\AA/min Eu flux corresponding to 25 s. per EuO layer, it
is the limited oxygen supply from the MBE environment which
dictates the speed during the sustained growth, i.e., to 30 s per
EuO layer.

To elucidate further the growth process, we have measured the
thickness of the films using \textit{ex situ} XRR measurements.
Since EuO deteriorates rapidly under ambient conditions, the films
need to be capped. To this end, an aluminum layer with a thickness
of 20--40~\AA\ has been evaporated on top of the EuO. This
thickness turns out to be sufficient for the aluminum to be a good
protective overlayer as will be discussed later. Figure~10
exhibits the XRR profile of the EuO film which was grown at an
oxygen pressure of 4$\times$10$^{-8}$ mbar for 200 min. The
corresponding RHEED and LEED patterns of the EuO film are
displayed in Fig.~8(a). From the period of interference fringes,
we deduce that the thickness of the EuO film is about 350~\AA.

Thicknesses of the other films are also determined from their XRR
profiles. The results are displayed in Fig.~11, where we plot the
thickness against the product of oxygen pressure and total
deposition time. For oxygen pressure up to $12\times10^{-8}$ mbar
we can observe a clear and direct linear relationship between
them, strongly suggesting that the thickness is determined by the
amount of oxygen incorporated. In other words, the growth is
limited by the amount of oxygen made available. This in turn means
that the Eu flux rate is higher than necessary and that the excess
Eu must be re--evaporated into the vacuum. Figure~11 essentially
confirms the distillation process needed to maintain good control
of the stoichiometry as reported in our earlier studies by
Steeneken \textit{et~al}.\cite{steeneken02a,steeneken02b} and
Tjeng \textit{et~al}.\cite{ott06a}

\begin{figure}[t]
\includegraphics*[scale = 0.3] {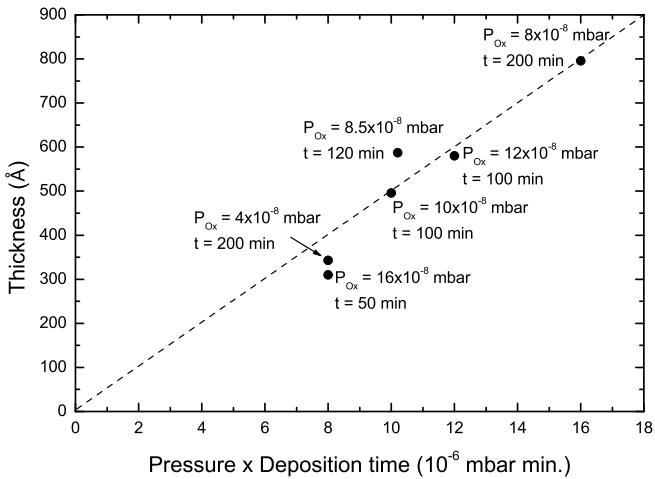}
\caption{\label{Fig_11} EuO film thickness, as determined from XRR
measurements, versus the product of oxygen pressure and total
deposition time.}
\end{figure}

\begin{figure}[t]
\includegraphics*[scale = 0.3] {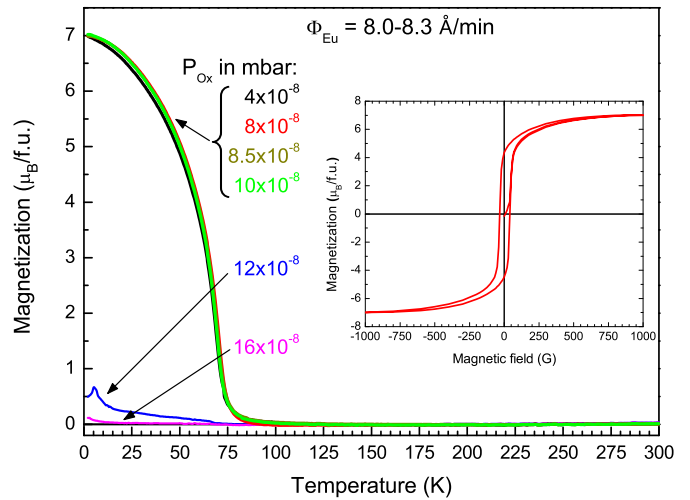}
\caption{\label{Fig_12} (Color online) Temperature dependence of
the magnetization of epitaxial EuO films on YSZ (001) grown at
$400^{\circ}$C with 8.0--8.3~\AA/min Eu flux rates under various
oxygen pressures as indicated. The small magnetization
contribution from the substrate has been subtracted. The applied
magnetic field was 1000 G. The inset shows the field dependence of
the magnetization of epitaxial EuO on YSZ (001) at 5~K. The film
was grown at $400^{\circ}$C with a 8.2~\AA/min Eu flux rate and a
$8\times10^{-8}$ mbar oxygen pressure.}
\end{figure}

We now investigate to what extent the growth conditions affect the
magnetic properties of the EuO films using a SQUID magnetometer.
The results are shown in Fig.~12. The films grown with 4, 8, 8.5,
and 10 $\times10^{-8}$ mbar oxygen pressures all have a Curie
temperature of 69 K with a magnetic moment of 7~$\mu_B$/f.u. as
expected for a $4f^7$ system. The inset shows the field dependence
of the magnetization at 5~K for the film grown with a
$8\times10^{-8}$ mbar oxygen pressure. Here one can observe a
hysteresis behavior with a saturation magnetization of 7$\mu_B$.
These results are in agreement with the RHEED and LEED results as
displayed in Figs. 8(a)--8(c), in the sense that the proper
ferromagnetic properties are always maintained as long as good
epitaxial growth is also achieved. On the other hand, films grown
with 12 and $16\times10^{-8}$ mbar have completely lost their
ferromagnetic properties. It is remarkable that exceeding the
$10\times10^{-8}$ mbar value just a little bit causes such a
dramatic change. This very abrupt change is also consistent with
the RHEED and LEED results as displayed in Fig.~8~(d).
Considerable film roughness starts to develop for oxygen pressures
higher than $10\times10^{-8}$ mbar. We would like to infer that
having only a $T_C$ of about 69~K is not a sufficient
characteristic to conclude that the film is homogeneous and
stoichiometric. One also needs to establish that the film has a
full saturation magnetization of 7$\mu_B$. The measurement of the
magnetic properties can therefore serve as a critical test for the
growth conditions and in particular, the oxygen stoichiometry of
the EuO films.

\begin{figure}[t]
\includegraphics*[scale = 0.3] {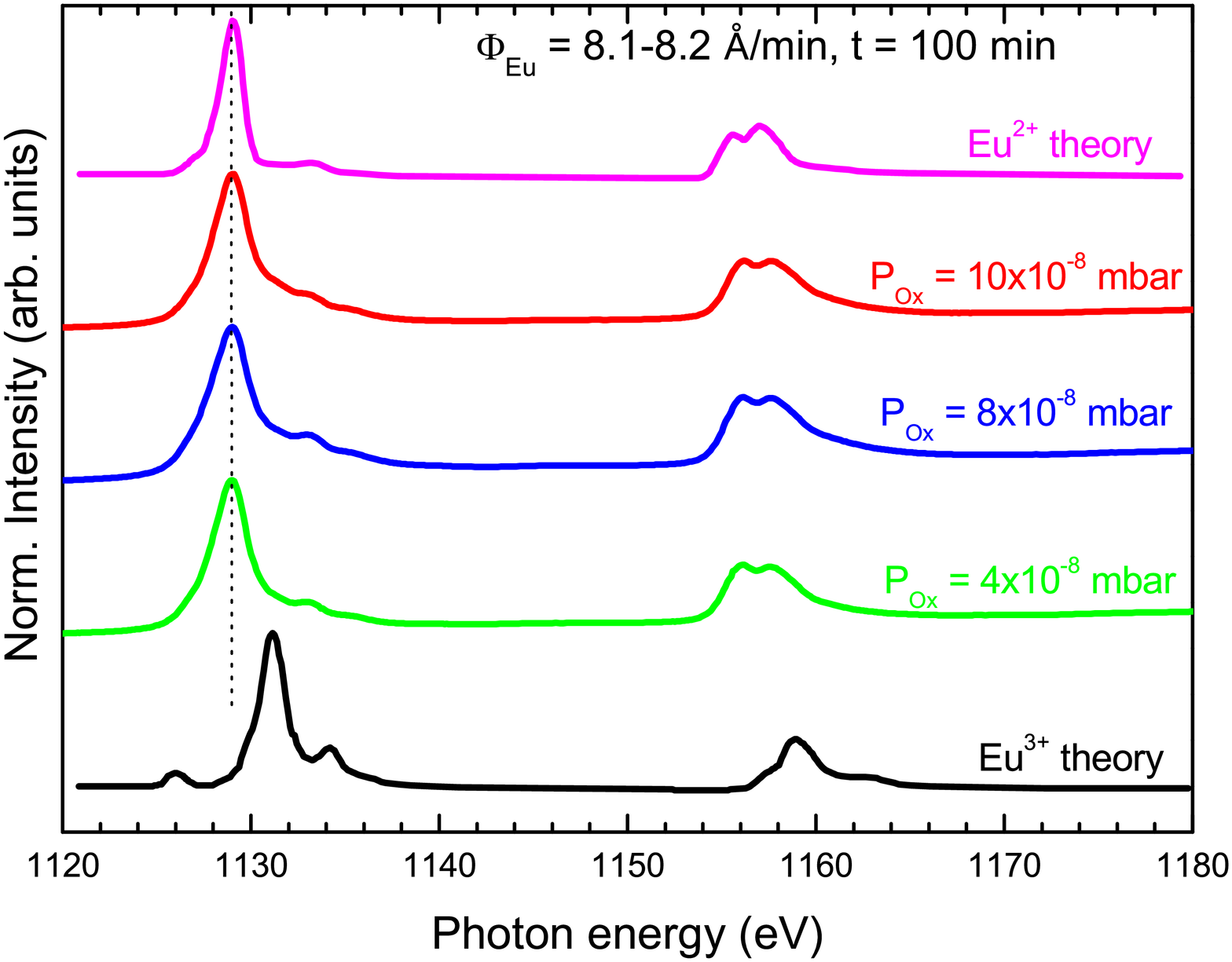}
\caption{\label{Fig_13} (Color online) Eu $M_{4,5}$ ($3d
\rightarrow4f$) XAS spectra of EuO films grown epitaxially on YSZ
(001) at $400^{\circ}$C with 8.1--8.2~\AA/min Eu flux rates under
various oxygen pressures as indicated. The films are capped with a
20--40~\AA~aluminum overlayer. Theoretical spectra of Eu$^{2+}$
and Eu$^{3+}$ are also shown, retrieved from Ref.~[46].}
\end{figure}

We have also performed \textit{ex situ} soft XAS measurements at
the Eu $M_{4,5}$ edges to examine the integrity of the EuO films
after capping with the aluminum overlayer. Figure~13 depicts the
XAS spectra together with the theoretical spectra for Eu$^{2+}$
(top) and Eu$^{3+}$ (bottom).\cite{thole85a,goedkoop88a} It is
clear that the Eu spectra are very similar to the theoretical
spectrum for Eu$^{2+}$, meaning that the EuO films with 4, 8, and
10 $\times10^{-8}$ mbar oxygen pressures are completely free from
Eu$^{3+}$ species. This in turn implies that an aluminum overlayer
as thin as 20--40~\AA\ works well to protect the EuO films against
air, contrary to the claims made elsewhere that one needs very
thick capping
layers.\cite{iwata00a,iwata00b,lettieri03a,matsumoto04a,holroyd04a,santos04a,negusse06a,schmehl07a,laan08a,ulbricht08a}
We attribute this to the fact that the epitaxial growth of EuO on
YSZ (001) yields such a smooth film so that a very thin aluminum
film is sufficient to make a closed capping overlayer. The
smoothness of the films as well as the complete absence of
Eu$^{3+}$ impurities forms a good starting point for the
fabrication of well--defined interfaces with other metals or oxide
materials, thereby opening up new opportunities to study or even
generate new phenomena related to interface physics.

We conclude that the critical oxygen pressure is around
10--12$\times10^{-8}$ mbar for a 8.0--8.3 \AA/min Eu flux rate.
Only below this pressure one has the distillation process taking
place so that good epitaxial growth can be achieved with the
proper stoichiometry and ferromagnetic properties. Apparently,
layer--by--layer growth can be obtained only if one is close to,
but not exceeding, the critical pressure.

\section{Epitaxial growth of E\lowercase{u}O on M\lowercase{g}O}
\begin{figure} [b]
\includegraphics*[scale = 0.3] {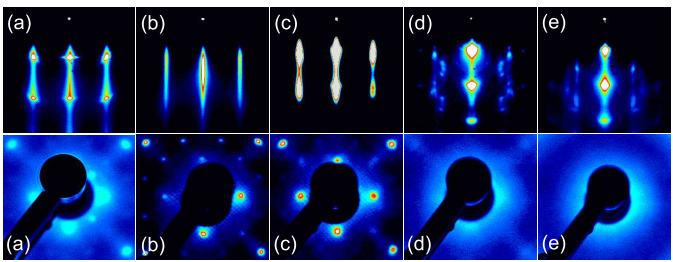}
\caption{\label{Fig_14} (Color online) Top panels: RHEED
photographs of EuO films on MgO (001) grown at $400^{\circ}$C with
8.1--8.2~\AA/min Eu flux rates under oxygen pressures of -- from
left to right -- (a) 4, (b) 8, (c) 10, (d) 12, and (e)
$16\times10^{-8}$ mbar. The deposition times were all 100 min. The
RHEED electron energy was 20 keV with the beam incident along the
[100] direction. Bottom panels: corresponding LEED photographs.
The LEED electron energy was set at 360 eV.}
\end{figure}

Having understood the growth process of EuO on YSZ (001), and
having found the recipe to obtain excellent epitaxial growth of
EuO on YSZ (001), we now turn our attention to the preparation of
EuO on MgO (001). MgO as a substrate for EuO is interesting since
several studies\cite{sohma97a,iwata00b,steeneken02b,lee07a} have
reported good epitaxial growth despite the very large lattice
mismatch of about 20\%. The RHEED and LEED photographs of EuO
films grown on MgO using different oxygen pressures are displayed
in Fig.~14. EuO films with oxygen pressure below
10$\times$10$^{-8}$ mbar all show excellent epitaxial growth. The
relationship between the distances of the MgO and EuO streaks is
consistent with the ratio of their lattice constants of 1.22. On
the other hand, a higher oxygen pressure creates additional spots
in the RHEED image and even causes the LEED pattern to disappear.
Similar to the EuO on YSZ case, this indicates that
10--12$\times$10$^{-8}$ is the critical value for the oxygen
pressure above which one no longer gets crystalline and
stoichiometric EuO.

Our attempts to observe RHEED intensity oscillations of EuO films
grown on MgO (001) were yet unsuccessful. The RHEED streaks and
specular spot suddenly disappeared right after the deposition of
EuO has been initiated. In approximately 30 s, new streak lines
appear, whose spacing conforms to the EuO lattice parameter.
However, the specular spot was never recovered. A more detailed
growth study is required to determine the growth process of EuO on
MgO (001).

To investigate the quality of these EuO films in terms of their
magnetic properties, we have performed SQUID measurement for films
grown with oxygen pressures of 8, 10, and 12$\times$10$^{-8}$
mbar. The results are shown in Fig.~15. Similar to the EuO on YSZ
case, the films that were grown below 10$\times$10$^{-8}$ mbar
have a Curie temperature of 69~K with a magnetic moment close to
7$\mu_B$/f.u. Conversely, the film grown under a higher oxygen
pressure has completely lost its ferromagnetic character. The
small peak at roughly 5~K indicates the typical antiferromagnetic
ordering temperature of Eu$_3$O$_4$, meaning that the dramatic
loss of the ferromagnetism for the growth using slightly above the
critical oxygen pressure is due to the presence of Eu$^{3+}$
species.

\begin{figure}[t]
\includegraphics*[scale = 0.3] {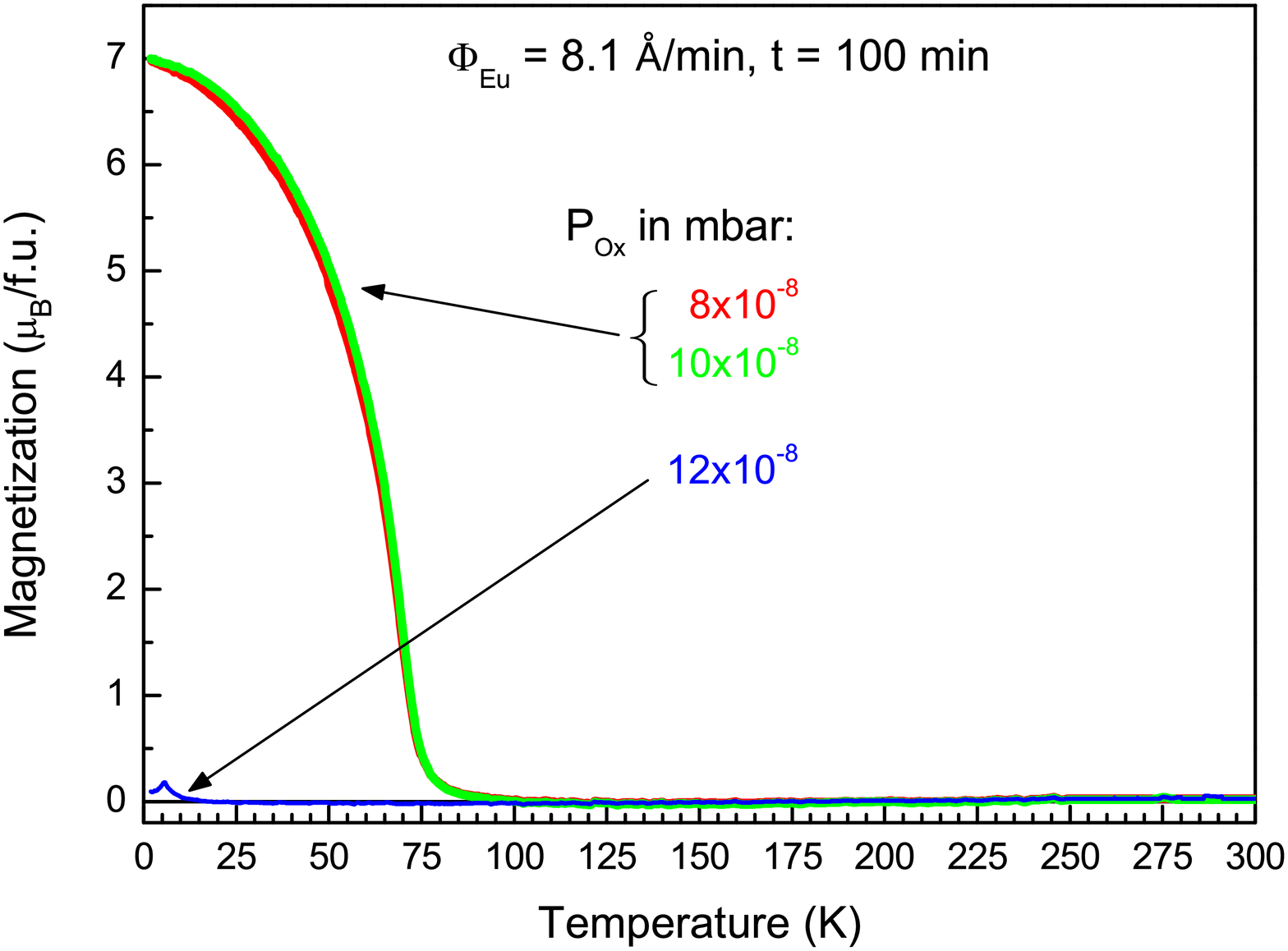}
\caption{\label{Fig_15} (Color online) Temperature dependence of
the magnetization of epitaxial EuO films on MgO (001) grown at
$400^{\circ}$C with a 8.1~\AA/min Eu flux rate under various
oxygen pressures as indicated. The small magnetization
contribution from the substrate has been subtracted. The applied
magnetic field was 1000 G.}
\end{figure}

\section{Conclusion}
We have successfully grown epitaxial and highly stoichiometric EuO
films on YSZ (001). The initial stages of growth involve the
limited supply of oxygen from the YSZ substrate, but the EuO
stoichiometry can still be well maintained. We have also observed
RHEED intensity oscillations during the sustained stages of
growth, which demonstrate that the layer--by--layer growth mode
can be achieved for EuO films on YSZ (001). The EuO films were
sufficiently smooth so that capping with a thin layer of aluminum
enabled \textit{ex situ} experiments free from trivalent Eu
species. The excellent epitaxial growth of EuO films is always
accompanied by equally excellent ferromagnetic properties: a
saturation magnetization of 7$\mu_B$ and a $T_C$ of 69 K. We have
also confirmed that the use of the Eu--distillation process during
the MBE--assisted growth enables the consistent achievement of
stoichiometry. The layer--by--layer growth, the smoothness of the
film, and the excellent stoichiometry provide an excellent basis
for the fabrication of well--defined interfaces for further
studies.

\section{Acknowledgments}
We would like to thank Lucie Hamdan for her skillful technical and
organizational assistance in preparing the experiment. We
acknowledge the NSRRC staff for providing us with beam time. The
research in Cologne is supported by the Deutsche
Forschungsgemeinschaft through SFB 608.

\end{document}